\documentclass[11pt]{article}

\usepackage[final]{acl}

\usepackage{times}
\usepackage{latexsym}
\usepackage[T1]{fontenc}
\usepackage[utf8]{inputenc}
\usepackage{microtype}
\usepackage{graphicx}
\usepackage{fvextra}
\usepackage{caption}

\title{Extracting Published Discussions of 13,616 Galaxies with Language Models}
\author{Nolan Koblischke \and Jo Bovy \\
  David A. Dunlap Department of Astronomy and Astrophysics \\
  Dunlap Institute for Astronomy \& Astrophysics \\
  University of Toronto, 50 St. George Street, Toronto, ON, M5S 3H4, Canada \\
  \texttt{nolan.koblischke@mail.utoronto.ca}}

\begin{document}
\maketitle

\begin{abstract}
Astronomers often have difficulty determining whether an object, such as a star or galaxy of interest, has been discussed in the literature. Existing tools enable object identification and linking in papers but do not record whether an object was genuinely discussed in a paper or just appears in a table or catalog. We address this by using language model agents to identify galaxies discussed in more than two sentences across a corpus of 156,006 arXiv astro-ph papers and extract those discussions. We extract 44,332 discussions of individual galaxies containing 85,314 quotes from 18,641 papers. With sky coordinates for 98.2\% of discussions, we group them into 13,616 galaxies. We bring these results together into GalaxyWiki, where users can find literature discussing galaxies, read what was said about them, and follow links to the source papers. The code, dataset, and webpage are available at \url{https://github.com/NolanKoblischke/galaxywiki-paper}.
\end{abstract}

\begin{figure*}
    \centering
    \includegraphics[width=1\linewidth]{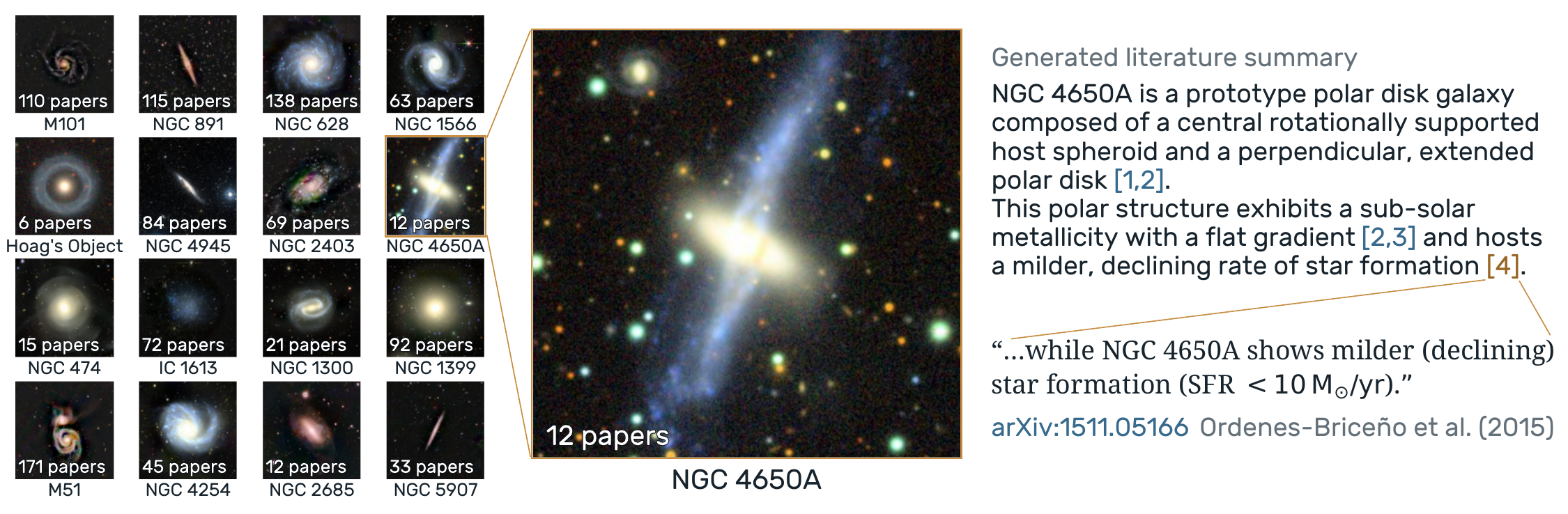}
    \caption{GalaxyWiki brings together published discussions of individual galaxies. \textit{Left:} Selected galaxies and the number of papers discussing each galaxy in our dataset. \textit{Right:} The two-sentence summary of the literature discussions of NGC 4650A~\citep{2010AIPC.1240..379I,2014arXiv1402.7325I,2011A&A...531A..21S,2016A&A...585A.156O}, along with a citation, extracted quote, and source paper.}
    \label{fig:header}
\end{figure*}

\section{Introduction}

The NASA/IPAC Extragalactic Database (NED) combines identifiers, physical measurements, and literature information for galaxies and other extragalactic objects~\citep{2020ASPC..527....3M}. Similarly, SIMBAD covers galaxies and stars, providing identifiers, basic data, selected measurements, and bibliographic references~\citep{2000A&AS..143....9W}. NED and SIMBAD also provide catalog-level object classifications like whether a galaxy is a spiral or an elliptical~\citep{2022MNRAS.515..807K}. Here we seek a complementary resource: a collected account of how individual galaxies have been interpreted scientifically, grounded in extended quotations from the literature.

Astronomers searching for previous discussions of a galaxy can find the object bibliographies provided by these databases~\citep{2020ASPC..527....3M,2000A&AS..143....9W}. However, an object-to-paper link alone does not distinguish an extended discussion from an appearance in a catalog. Readers may therefore need to inspect the linked papers to determine what has actually been said about the galaxy.

Wikipedia provides this kind of scientific interpretation, although its coverage is limited. We count 3,379 English Wikipedia article pages that use the \textit{galaxy} infobox, a proxy for its coverage of galaxies.\footnote{\url{https://linkcount.toolforge.org/?project=en.wikipedia.org\&page=Template\%3AInfobox+galaxy}} However, these pages might not provide a complete account of the literature about a given galaxy. Galaxy Zoo offers another setting for discussion, allowing volunteers to classify galaxies~\citep{2008MNRAS.389.1179L} and discuss them on forums that bring together citizen scientists and experts~\citep{2011ASPC..443..150M}.

To simplify this search, we aim to extract every galaxy discussed in more than two sentences in our corpus of arXiv papers, along with quotes. We bring these discussions together in a database that captures scientific interpretations. Astronomers can then immediately check whether a galaxy has been discussed in the extracted literature, read what was said, and identify the papers containing those discussions, as illustrated in Figure~\ref{fig:header}.

\subsection{Related Work}
Recent work has used language models to extract information from scientific papers and organize it into resources that are easier to search and read~\citep{2023arXiv231207559L,2024arXiv240913740S}. WikiCrow generates cited articles about human genes from the research literature using PaperQA2 agents that search papers and synthesize their findings~\citep{2024arXiv240913740S}. In astronomy, \textit{Lodestone} uses language models to bring together findings from multiple papers in cited encyclopedia articles~\citep{iyer2026lodestone}. \citet{2024RAA....24f5012S} extracts celestial object identifiers and telescope names from journal articles. \citet{2026RAA....26c5011W} constructs a galaxy knowledge graph from literature abstracts, while \citet{2026arXiv260214335L} resolves object mentions to SIMBAD identifiers and links them to scientific concepts. With GalaxyWiki, we focus on the literature discussions of individual galaxies, rather than just their properties or identifiers.

\section{Methods}
\paragraph{Source corpus}
We use preprocessed arXiv paper text from the MINT-1T dataset~\citep{2024arXiv240611271A} that is created by parsing the source LaTeX into text~\citep{2024arXiv240611271A}. Specifically, we use \texttt{minty-astro-ph},\footnote{\url{https://huggingface.co/datasets/Smith42/minty-astro-ph}} a subset of MINT-1T focused on astro-ph papers. This contains 156,006 astro-ph papers from June 1993 to May 2024. For comparison, we count 299,824 astro-ph postings in the same time period according to the Kaggle arXiv metadata dataset.\footnote{\url{https://www.kaggle.com/datasets/Cornell-University/arxiv}} This incomplete coverage appears largely attributable to the MINT-1T figure-extraction requirements, since the original corpus was designed to pair text with images. In future work, we will attempt to construct a complete astro-ph corpus. As a first processing step, we apply a filter that retains only papers containing the standalone word ``galaxy'' or ``galaxies,'' leaving 87,899 papers.

\paragraph{Galaxy mention extraction}
We process each paper with the GPT-5.5\footnote{\url{https://developers.openai.com/api/docs/models/gpt-5.5}} language model in the open-source Codex\footnote{\url{https://github.com/openai/codex}} agent harness, and task it with identifying mentioned galaxies and extracting quotes. In the Codex harness, the language model has access to terminal-based tools that allow it to copy quotes verbatim and selectively read the paper. For each galaxy identified, we request one to three quotes totaling three to eight sentences, and ask the language model to generate a 20-to-40-word summary supported by those quotes. We also ask it to choose up to five relevant topics drawn from a fixed subset of the Unified Astronomy Thesaurus (UAT; \citealp{2018ApJS..236...24F}), enabling thesaurus-driven search in our GalaxyWiki. We further task the language model with extracting the galaxy's sky coordinates when explicitly provided in the paper, which is required for associating galaxies across multiple papers. We explicitly instruct the language model to ignore mentions of the Milky Way Galaxy and objects within it and instruct it to extract at most 30 galaxies per paper.

To reduce the setup and take-down overhead of the agents, we assign each agent a batch of 10 papers. We run 20 concurrent agents for approximately 71 hours with token usage corresponding to approximately \$12,600 USD at standard API prices.\footnote{\url{https://developers.openai.com/api/docs/pricing}} We extract 44,332 discussions of individual galaxies made of 85,314 quotes contained in 18,641 unique papers.

\paragraph{Coordinate resolution for grouping discussions}
Access to the position of the discussed galaxy on the sky allows us to associate discussions across papers and with other catalogs. However, our paper extraction step only captures coordinate information for 3,669 mentions, or 8.3\%.

Instead of relying on the coordinates from the extraction step, and to vastly improve coordinate coverage, we use the extracted galaxy identifiers if they follow standard formats such as the NGC, SDSS, and 2MASS identifiers~\citep{2000AJ....120.1579Y,2006AJ....131.1163S} to match them to an entry on the NED database~\citep{2017IAUS..325..379M}. We programmatically query the NED API using galaxy identifiers that match these standard formats. Using this method, we successfully retrieve coordinates for 23,770 out of 25,206 attempted identifier queries, which populates coordinates for 53.6\% of mentions.

For the remaining mentions, we task GPT-5.5 agents with finding coordinates by first instructing them to search the original paper, then query the NED API using any compatible galaxy identifier from the paper, not necessarily the formats we attempted. As a fallback, we provide internet access to allow the agents to follow links and references in the paper and retrieve resources such as VizieR catalogs published by the authors~\citep{2000A&AS..143...23O}. We record the internet resources used so that we can audit the coordinate search and we ask the agents to report ``unresolved'' if they determine the task is impossible.

With agents, we find coordinates for 6,701 records in the source papers, 11,503 through NED, and 1,684 from external internet sources. These results provide coordinates for 19,888 records, or 44.9\% of all mentions, leaving 674 records unresolved. Across the programmatic and agentic methods, coordinates are available for 43,546 of 44,332 mentions, or 98.2\%.

We then group the 43,546 located galaxy mentions across papers using these sky coordinates. We link positions separated by less than 1$^{\prime\prime}$ and include neighbors of neighbors. The procedure yields 13,616 entries, each of which we interpret as an individual galaxy discussed in the source papers. However, this relies on the choice of 1$^{\prime\prime}$ and grouping by sky position does not ensure that these entries correspond to distinct galaxies.

\paragraph{GalaxyWiki}
In addition to releasing a dataset, we built a web-based wiki intended for astronomers and the general public. For each galaxy page, we generate a two-sentence summary with citations from the available extracted quotes and paper-level summaries using Gemini 3.5 Flash.\footnote{\url{https://ai.google.dev/gemini-api/docs/models/gemini-3.5-flash}} We programmatically verify that the source quotes and papers used for the summaries are in our dataset. Alongside this summary, each page lists the papers, their summaries, and their quotes, with direct links to the quoted passages where available through the arXiv HTML interface. We allow users to search for galaxies by sky position, UAT topic, name, or text description. To enhance user experience, we retrieve available preview images from the DESI Legacy Imaging Surveys DR10~\citep{2019AJ....157..168D}, Pan-STARRS~\citep{2016arXiv161205560C}, and Digitized Sky Survey 2~\citep{2000ASPC..216..145M} using the coordinates.

\section{Results}
The extraction yields 44,332 galaxy mention records containing 85,314 quotes from 18,641 papers. The 13,616 grouped entries include 4,187 with multiple mentions and 9,429 (69.2\%) with a single mention. Among entries with multiple mentions, the median is three mentions per entry. Every page in the wiki includes a preview image from Legacy Survey DR10 (9,545 pages), Pan-STARRS (3,831 pages), or DSS2 (240 pages).

To assess whether agents accurately copied quotations, we compare each extracted quote with the source paper text. We find character-for-character matches for 66,362 of the 85,314 quotes (77.8\%). Normalizing Unicode, spacing, and punctuation yields another 15,971 matches, bringing the total to 82,333 (96.5\%). A further 34 match after ignoring case and removing all characters other than letters and numbers. Finally, we find that 2,826 can be split at sentence punctuation into fragments that each match the source after Unicode and whitespace normalization, indicating that agents often combined non-contiguous excerpts into a single quotation. However, textual matching alone does not verify correct attribution or preservation of the source's meaning.

We manually review 120 distinct galaxy mention records. We first review 60 randomly selected records with coordinates, focusing on coordinate verification. We find support for the intended galaxy identification and correct quotations in 58 cases. Two records contained quotes that provided little substantive discussion or described a different object. In three otherwise correctly identified records, the galaxy identifiers were truncated or the paper-supplied positional coordinates were rounded, which reduced the precision of the extracted coordinates, with the largest discrepancy reaching 12.2$^{\prime\prime}$ because an agent decoded a shortened identifier that contained coordinate information. Such errors can separate discussions of the same galaxy into different entries.

We then review another 60 records drawn at random, this time without restricting coordinate availability. During review, all 60 records identify an individual galaxy and have quotes that contain at least three distinct substantive sentences about that galaxy. Two records have passages that only discuss the galaxy jointly with other galaxies, though we determine that these passages still provide useful information for the GalaxyWiki. Four records contain source text in which the MINT-1T parsing pipeline deleted or corrupted LaTeX symbols. This omits details that could be useful to understand the discussed galaxy. Despite these issues, the review supports the conclusion that the pipeline achieves its primary goal, correctly identifying an individual galaxy and extracting its substantive discussion in 118 of 120 cases.

\section{Discussion}
Our GalaxyWiki is a unique dataset for the field, and has already been used for 1) determining if galaxies discovered by an anomaly detection method have been previously discussed in the literature~\citep{cheng2026interesting} and 2) enhancing descriptions of galaxy images for training an astronomy-specific vision-language model~\citep{habib2026astrobridge}.

The GalaxyWiki dataset provides a literature-selected sample of galaxies that have received extended scientific discussion and are therefore of interest to the community. Therefore, one could use the catalog to guide similarity and semantic searches for galaxies with related features~\citep{2021arXiv211013151S,2026ApJ..1005..170K} and support the development of methods for detecting anomalies or identifying galaxies of scientific interest~\citep{2021A&C....3600481L,2024MNRAS.529..732E}. The database could also support literature surveys in astronomy, complementing semantic retrieval and corpus-wide concept extraction~\citep{2024ApJS..275...38I,2025arXiv251112353T}. We can now track which galaxies are discussed and cited most often and how these patterns vary across subfields and years.

The same literature-processing methodology could, in principle, extend beyond galaxies to stars, supernovae, exoplanets, and even theories of dark matter. It could also support similar databases beyond astronomy, drawing on bioRxiv for genes or on ChemRxiv for materials, building on precedents for gene-focused literature synthesis~\citep{2024arXiv240913740S} and structured extraction of materials information~\citep{2024NatCo..15.1418D}.

\section{Conclusion}
We create GalaxyWiki, a database that brings together scientific interpretations of galaxies and discussions of what makes them noteworthy. It allows astronomers and the general public to check whether a galaxy has been discussed in the extracted literature, read those discussions, and continue to the relevant papers. GalaxyWiki illustrates how language models can make previously impractical scientific projects feasible by performing parallelizable and verifiable tasks. Future work will expand the corpus and evaluate mention identification, coordinate resolution, and grouping to build a more complete collection.

\section*{Limitations}
Even though we can verify that the extracted quotes are in the original papers, we cannot ensure that we have retrieved the mentions of every galaxy in a given paper. It is likely that the language model misses the mentions of some galaxies during the extraction step. In fact, our setup guarantees that some galaxies have been missed since we set a limit of 30 galaxies per paper, which is hit for 39 papers. Notably, in one paper the language model recorded 250 galaxies despite this prompted limit, which we inspect and find minor errors in some extracted sentence counts.

Our coverage is also limited by the source corpus, because we use only the astro-ph subset available through MINT-1T~\citep{2024arXiv240611271A}. Papers without an arXiv version are also absent. Expanding the corpus in future work would help produce a more complete dataset.

In our programmatic NED resolution, a pattern intended to capture Abell designations~\citep{1958ApJS....3..211A} also matched local identifiers like ``Galaxy A1'', producing false positives. After identifying the issue, we invalidated 112 of the 156 rows resolved with this pattern.

\section*{Acknowledgments}
This work was enabled by credits provided through OpenAI's ChatGPT Futures award. This work also used API credits provided by Google through its Gemini Academic Program. Author NK also acknowledges the support of the Natural Sciences and Engineering Research Council of Canada (NSERC) [funding reference number 721697856] and the Data Sciences Institute at the University of Toronto through grant number DSI-DSFY4R1P01. We thank Gabriel Pfaffman and James Garland for valuable discussions and helpful feedback that improved this work. This work used the NASA/IPAC Extragalactic Database and the SIMBAD database.
NK and JB acknowledge support from the Natural Sciences and Engineering Research Council of Canada (grant RGPIN-2026-04837).
\bibliography{custom}

\end{document}